\documentclass[grl]{agutex}

\usepackage{url}
\usepackage{graphicx}
\usepackage{lineno}

\authorrunninghead{SVALGAARD}

\titlerunninghead{CALIBRATION OF SUNSPOT NUMBERS}

\begin{document}

\title{Objective Calibration of Sunspot Numbers}

\authors{L. Svalgaard\altaffilmark{1}}
\altaffiltext{1}{HEPL, Via Ortega, Stanford University, Stanford, CA 94305, USA.}
\authoraddr{L. Svalgaard, HEPL, Via Ortega, Stanford University, Stanford, CA 94305, USA.
(leif@leif.org)}

\begin{abstract}
\cite{wald71} found a very tight relationship between the F10.7 solar radio flux and the sunspot number and suggested using the flux for an objective calibration of the sunspot number. He suggested that if this relationship changed later on, the sunspot number should be re-calibrated, assuming that the calibration must have drifted with time. I repeat his analysis using data up to the present and it is, indeed, clear that the relationship has changed significantly. This could be due to a drift of the calibration or to a secular change in the visibility of sunspots, or both.
\end{abstract}

\begin{article}
\section{The Sunspot Number Scale}
The sunspot number is the solar index most frequently used in the study of long-term solar activity variations and, even more so, in solar-terrestrial relation studies. The relative sunspot number was defined by Rudolf Wolf \citep{wolf56} as $R=k(10g+s)$ where $g$ is the number of sunspot groups, $s$ is the total number of `spots' in all the groups on the visible disk, and $k$ is a scale factor to bring the number on to Wolf's scale (thus $k=1$ for Wolf himself). It would seem that $g$ and $s$ should be uniquely determined simply by counting and that a $k$-factor would not be necessary. However, different observers - even using the same instrument - may differ in how they overcome variable seeing and arrive at different numbers of groups. Even more so for the number of spots, where the very definition of what should be counted as a spot may vary from observer to observer. Issues here are the distinction between spots with penumbrae and pores without, the treatment of spots that touch each other, different weighting according to size, whether to count all spots or only the larger ones, observer acuity and Snellen Ratio, etc. Different observers have different answers to, preferences of, and opinions about these issues. With an appropriate $k$-factor their observed counts are, presumably, reduceable to the Wolf-scale. There is some confusion as to the precise meaning of the reduction factor. Strictly speaking, it should only apply to the Standard Instrument: 8 cm refractor at magnification 64. To compensate for a larger or smaller telescope an additional factor should be employed. In practice this is too cumbersome and the additional factor is folded into the $k$-factor. For a detailed discussion of  these issues see \cite{scha93} and \cite{hoss01}.

\section{The Solar Microwave Flux}
The F10.7 cm (2.8 GHz) solar index, introduced by Covington in 1947 is generally viewed as an excellent index of solar activity \citep{tapp94}. The basic minimum level of emission around 67 sfu (solar flux unit =  $10^{-22}$ W m$^{-2}$ Hz$^{-1}$) is presumed to come from the quiet background Sun. An $S$-component due to solar activity on time scales longer than those of flares is fashioned into the F10.7 index measured and maintained by The Solar Radio Monitoring Programme operated jointly by the National Research Council and Natural Resources Canada (\url{http://www.spaceweather.gc.ca/sx-eng.php}). Routine observations at 1.0, 2.0, 3.75, and 9.4 GHz straddling the 2.8 GHz frequency of the Canadian series have been made in Japan since the 1950s (\url{ftp://solar.nro.nao.ac.jp/pub/norp/data}). The two microwave datasets [suitably scaled] compare favorably with one another and testify to the stability (to within the accuracy of the measurements) of the calibration of both \citep{sval10}.  

\section{The Calibration}
\begin{figure}
\centerline{\includegraphics[scale=0.75]{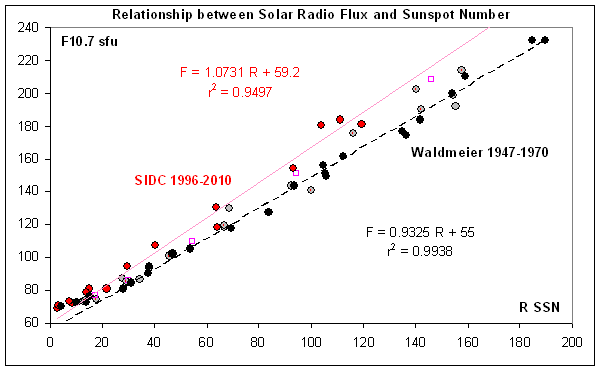}} 
\caption{Relationship between yearly means of solar 10.7 cm radio flux and sunspot number. The black dots and dashed regression line [for sunspot number greater than 25] are from \cite{wald71}. Gray dots are for 1971-1995 (with a small red center dot after SIDC took over in 1981). Red squares are for 1991-1995, and red dots for 1996-2010, {\it i.e.} solar cycles 23 and 24. 
} \label{F-Relation} 
\end{figure}

In a short 1971 paper, Max Waldmeier \citep{wald71} pointed out that ``the Z\"{u}rich standard scale [of the relative sunspot numbers] has never been calibrated in an objective way". He went on to note that the close correlation between monthly, and especially yearly, means of the solar microwave emission at 10.7 cm wavelength and the sunspot numbers yields a possibility of an objective calibration of the scale of the relative sunspot numbers. Figure~\ref{F-Relation} shows the tight relationship [linear for sunspot number greater than 25] deduced by Waldmeier for the interval 1947-1970 (black dots). He remarks that ``As long as this relation holds, the Z\"urich series of sunspot-numbers may be considered to be homogeneous. If this relation should be subject to changes in the time to come, then the reduction factor used hitherto ought to be changed in such a way that the old R-F relation is reestablished".

\begin{figure}
centerline{\includegraphics[scale=0.75]{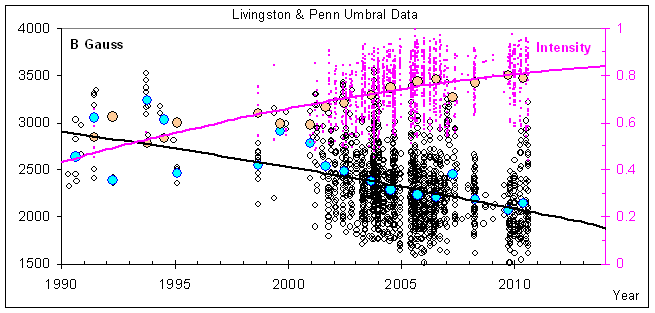}} 
\caption{Measurements of the magnetic field [in Gauss, black circles] in the darkest part of 1565 sunspot umbrae (years 1990-2010) and of the umbral intensity [small pink dots] relative to nearby non-spot photosphere (Livingston, 2010, personal communication). The larger blue and pink circles show yearly mean values.
} \label{F-Umbral} 
\end{figure}

Figure~\ref{F-Relation} also shows the relation since 1996 derived from the International Sunspot Number as determined by SIDC (red dots, \url{http://sidc.oma.be/DATA/yearssn.dat}). The data for the intervening interval 1971-1995 are shown as gray dots and open red squares. It is clear that the recent sunspot numbers no longer follow the relationship found by Waldmeier and that therefore, perhaps, ``the reduction factor used hitherto ought to be changed in such a way that the old R-F relation is reestablished". On the other hand it is also possible that the sunspot number as currently defined simply is no longer a suitable measure of solar activity, given the progressive discrepancy with the F10.7 flux. A similar conclusion was reached by \cite{sval10} and \cite{tapp10} based on monthly values. One can speculate that the reason for this is the recent reduced visibility of sunspots due to diminished contrast to the surrounding photosphere reported by \cite{penn06} on account of weaker magnetic field and increased temperature. Recent measurements appear to confirm those trends as seen in Figure~\ref{F-Umbral} [Livingston, personal communication]. Should such deviations from `normal' observed sunspot activity be substantiated in the near future, the question naturally arises whether [and when] they might have occurred in the past as well, {\it e.g.} during the Maunder Minimum, 1645-1715. 
   
\begin{acknowledgments}
I thank Bill Livingston for his measurements of umbral magnetic field and intensities. I acknowledge the use of  sunspot data from SIDC, RWC Belgium, World Data Center for the Sunspot Index, Royal Observatory of Belgium. And of F10.7 cm flux from National Research Council and Natural Resources, Canada. 
\end{acknowledgments}


\end{article}

\clearpage

\clearpage


\begin{thebibliography}{}
\bibitem[{\textit{Penn and Livingston}(2006)}]{penn06} Penn, M. J., and W. Livingston (2006), Temporal Changes in Sunspot Umbral Magnetic Fields and Temperatures, {\it Astrophys. Journal,} \textit{649(1)}, L45-L48, doi: 10.1068/508345.
\bibitem[{\textit{Hossfield}(2001)}]{hoss01} Hossfield, C. H. (2001), A history of the Zurich and American Relative Sunspot Number Indices, {\it J. Amer. Assoc. Var. Star Obs.,} \textit{30}, 48-53.
\bibitem[{\textit{Schaefer}(1993)}]{scha93} Schaefer, B. E. (1993), Visibility of sunspots, {\it Astrophys. Journal,} \textit{401}, 909-919.
\bibitem[{\textit{Svalgaard and Hudson}(2010)}]{sval10} Svalgaard, L., and H. S. Hudson (2010), The Solar Microwave Flux and the Sunspot Number, {\it SOHO-23: Understanding a Peculiar Solar Minimum, ASP conference series}, eds.: S. R. Cranmer, J. T. Hoeksema, and J. L. Kohl, \textit{428}, 325-328.
\bibitem[{\textit{Tapping and Charrois}(1994)}]{tapp94} Tapping, K. F., and D. P. Charrois (1994), Limits to the Accuracy of the 10.7 cm Flux, {\it Solar Phys.,} \textit{150}, 305-315.
\bibitem[{\textit{Tapping}(2010)}]{tapp10} Tapping, K. F., (2010), Properties of the sunspot Number and the 10.7 cm Solar Flux Activity Indices, their Interrelationship and Unusual Behaviour Since the Year 2000, {\it SORCE Meeting 2010, Keystone, CO.} \url{http://lasp.colorado.edu/sorce/news/2010ScienceMeeting/doc/Session6/6.03_Tapping_F10.7.pdf}.
\bibitem[{\textit{Waldmeier}(1971)}]{wald71} Waldmeier, M. (1971), An Objective Calibration of the Scale of Sunspot-Numbers, {\it Astron. Mitt. Eidg. Sternwarte Z\"{u}rich,} \textit{304}, pp10, [also available at \url{http://www.leif.org/research/W-CCCIV.pdf}].
\bibitem[{\textit{Wolf}(1856)}]{wolf56} Wolf, J. R. (1856), Mittheilungen \"uber der Sonnenflecken, {\it Astron. Mitt. Eidg. Sternwarte Z\"{u}rich,} \textit{2}, pp13.
\end{thebibliography}
\end{document}